\documentclass[letterpaper, 10 pt, conference]{ieeeconf}  

\IEEEoverridecommandlockouts                              

\usepackage{amsmath} 
\usepackage{amssymb}  
\usepackage{graphicx}
\usepackage{caption}
\usepackage{algorithm}
\usepackage{algorithmic}
\usepackage{subcaption}
\usepackage{array}
\usepackage{ragged2e}
\usepackage{rotating}  
\usepackage{booktabs} 
\usepackage{tabularx} 
\usepackage{caption}  

\title{\LARGE \bf
A Comprehensive Review of Propagation Models in Complex Networks: From
Deterministic to Deep Learning Approaches
}

\author{Bin Wu$^{1}$, Sifu Luo$^{2}$ and C. Steve Suh$^{1}$
\thanks{*\textbf{This paper is accepted for publication by Journal of Vibration Testing and System Dynamics and all copyright reserved}}
\thanks{$^{1}$The authors are with the Department of Mechanical Engineering, 
        Texas A\&M University, College Station, TX, 77840, USA
        {\tt\small $\{$wubin, ssuh$\}$@tamu.edu}}%
\thanks{$^{2}$Sifu Luo is with the Department of Physics \& Astronomy, 
        Texas A\&M University, College Station, TX, 77840, USA
        {\tt\small sifuluo@tamu.edu}}%
}

\begin{document}

\maketitle
\thispagestyle{empty}
\pagestyle{empty}

\begin{abstract}
    Understanding the mechanisms of propagation in complex networks is critical for various domains such as epidemiology, social media, communication networks, and multi-robot systems. This paper provides a comprehensive review of propagation models in complex networks, ranging from traditional deterministic models to advanced data-driven and deep learning approaches. 
We first discuss static and dynamic network structures, noting that static models offer foundational insights into network behavior, while dynamic models capture the time-evolving nature of real-world systems. Deterministic models, such as the SIR framework, provide clear mathematical formulations for describing the spread of information and viruses, but they often lack flexibility in dealing with real-world randomness. 
In contrast, stochastic models introduce randomness, making simulations of network behaviors more realistic, albeit at the expense of interpretability. Behavior-based models, including agent-based simulations, focus on individual decision-making processes, offering greater flexibility but requiring significant computational resources. Data-driven approaches leverage large datasets to adapt to changing network environments, improving accuracy in nonlinear and dynamic scenarios. These approaches can rely on the aforementioned models or be based on model-free machine learning methods.
We then explore supervised learning methods that require large amounts of labeled data, and unsupervised learning methods, which do not rely on labeled data. These two methods are the most mainstream approaches in machine learning. Building on this, we further investigate reinforcement learning, a newer learning paradigm that interacts with environments and does not require datasets.
Finally, we specifically discuss the application of graph neural networks (GNNs), which are closely aligned with network problems and have achieved revolutionary progress in modeling and optimizing propagation capabilities in large-scale and complex networks. The paper highlights key applications and challenges for each model type and emphasizes the growing role of hybrid and machine learning-based models in solving modern network propagation problems.
   
\end{abstract}

\begin{keywords}
Network Propagation, Machine Learning, Deterministic Model, Model-based Approaches, Model-free Approaches\end{keywords}

\section{Introduction}
The problem of propagation \cite{wen2012modeling, jiang2016identifying,chen2022information} in complex networks holds significant theoretical and practical importance in modern scientific research. With the rapid development of various complex networks such as social media \cite{jain2019stochastic, pierri2019false}, communication networks \cite{ren2018stochastic,hsu2020reinforcement}, and biological networks, understanding how information \cite{wu2018temporal,lu2022analytical}, viruses \cite{datilo2019review,choi2019unsupervised}, or influence spread \cite{yu2021self,shan2024diffusion} within these networks has become a key topic of study \cite{newman2003structure,strogatz2001exploring}. This not only helps us predict and control the spread of epidemics but also optimizes information dissemination strategies, enhancing the robustness and efficiency of networks.

This paper first provides an overview of the basic concepts and classifications of propagation problems in complex networks, clarifying the distinction between static and dynamic complex networks. Static complex networks have fixed node and edge structures, suitable for analyzing the static properties of networks. However, most real-world networks are dynamic, with nodes and edges constantly changing over time. Therefore, the concept of dynamic complex networks is introduced \cite{lin2020modeling,van2022dynamic,chen2022statistical,wu2018general}, offering a more accurate depiction of the evolutionary characteristics and dynamic behaviors of real networks.
Next, this paper explores in detail the modeling approaches used to capture propagation characteristics in dynamic complex networks. These methods are divided into model-based and model-free approaches. Model-based methods include deterministic models, stochastic models, behavior-based models, and data-driven models. These methods describe the propagation dynamics of network systems by establishing explicit mathematical or computational models. For example, the classic SIR model \cite{kermack1927contribution} and the random walk model \cite{kogias2009study} are widely used to simulate disease transmission and information diffusion.
Model-free approaches are primarily based on machine learning and deep learning techniques, including supervised learning, unsupervised learning, reinforcement learning, and graph neural networks. These methods do not rely on explicit system models but instead, automatically learn the propagation characteristics of networks from large amounts of data or interacting with the environment. In particular, graph neural networks \cite{wu2020comprehensive,zheng2022graph} excel at handling network-structured data, effectively capturing the complex relationships and dependencies between nodes.
Finally, this paper discusses the advantages and challenges of the aforementioned methods in practical applications. Although deterministic models have clear mathematical structures, they may be overly simplistic when dealing with real-world randomness and uncertainty. Stochastic models introduce random variables, allowing for more realistic simulation of propagation processes, but they also bring increased computational complexity and interpretive difficulty. Behavior-based models and data-driven models excel in flexibility and adaptability, but they face challenges related to parameter tuning and data quality. Machine learning methods have advantages in handling large-scale data and complex pattern recognition, but they are highly data-dependent or environment-dependent, and their models are often less interpretable.

Through the analysis and comparison of various methods, this paper aims to provide researchers with a comprehensive perspective, helping them choose the most suitable methods and tools for addressing propagation problems in complex networks. Future research directions may include the integration and improvement of methods, as well as in-depth exploration in different application domains.

\section{Overview of Propagation Problem in Complex Network}
When further discussing issues related to network propagation, we must clarify some basic concepts \cite{newman2003structure,strogatz2001exploring}. A static network is a fixed and unchanging network structure \cite{yue2024critical}, meaning that the nodes and edges in the network remain constant throughout the observation period. This type of network serves as the foundation for many studies in network theory, providing a simplified framework for analyzing the structural characteristics and basic behavior of networks without involving time factors. We can define it as a graph \( G = (V, E) \), where \( V \) is the set of nodes and \( E \subseteq \{(i, j) | i, j \in V\} \) is the set of edges. Depending on the nature of the edges, the network can be either directed or undirected. In a directed network, an edge \( (i, j) \) represents a one-way connection from node \( i \) to node \( j \); in an undirected network, the edges represent bidirectional or undirected connections between nodes. A typical example of a static complex network is a book co-citation network. In such a network, nodes represent books, and edges represent instances where two books are cited together by the same academic paper or other publication. This type of network is static because once established, the nodes (books) and edges (co-citation relationships) generally do not change.

\begin{figure*}[t]
    \centering
    \begin{minipage}{0.18\textwidth}
        \centering
        \includegraphics[width=\textwidth]{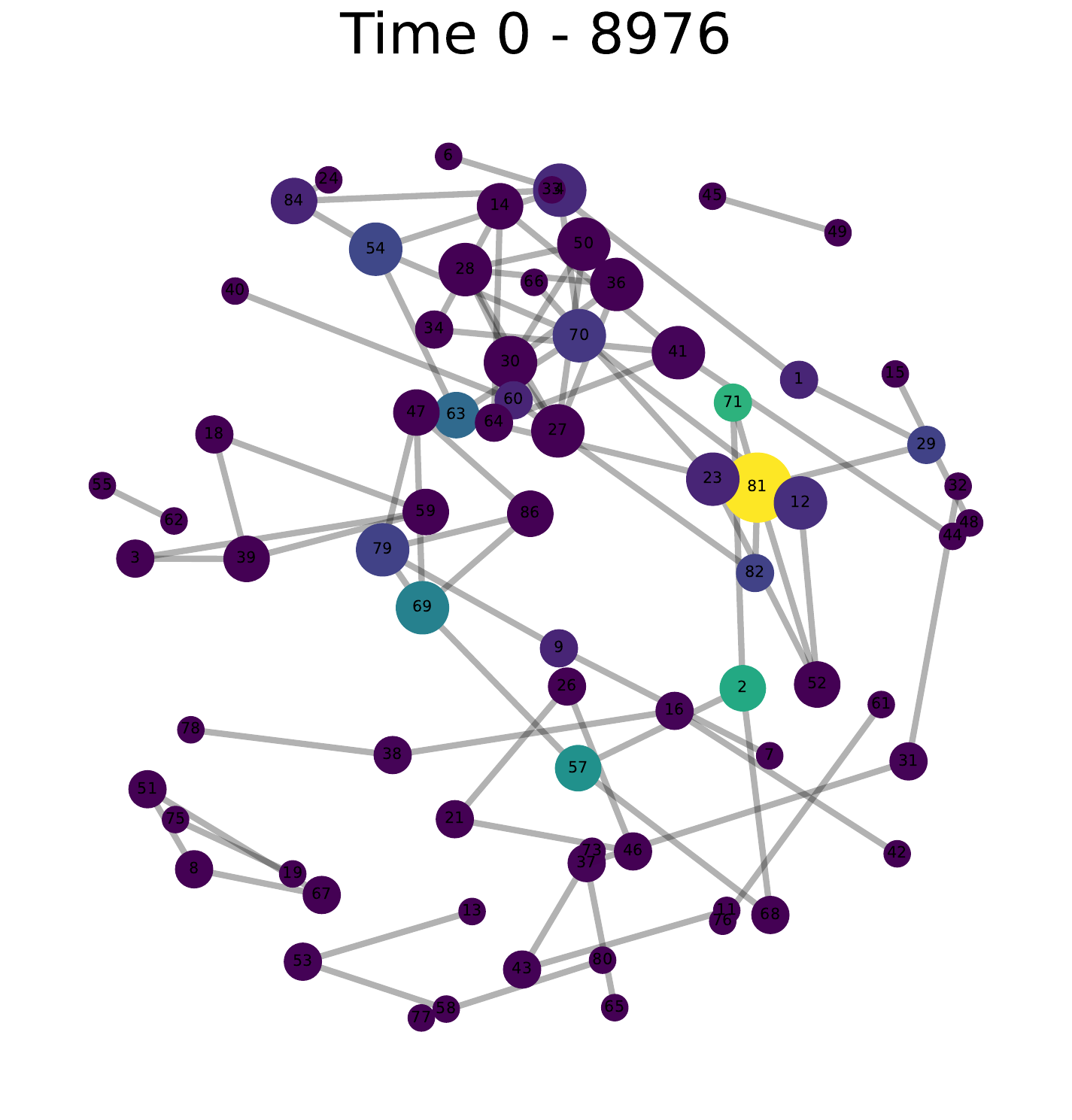}
    \end{minipage}
    \hfill
    \begin{minipage}{0.18\textwidth}
        \centering
        \includegraphics[width=\textwidth]{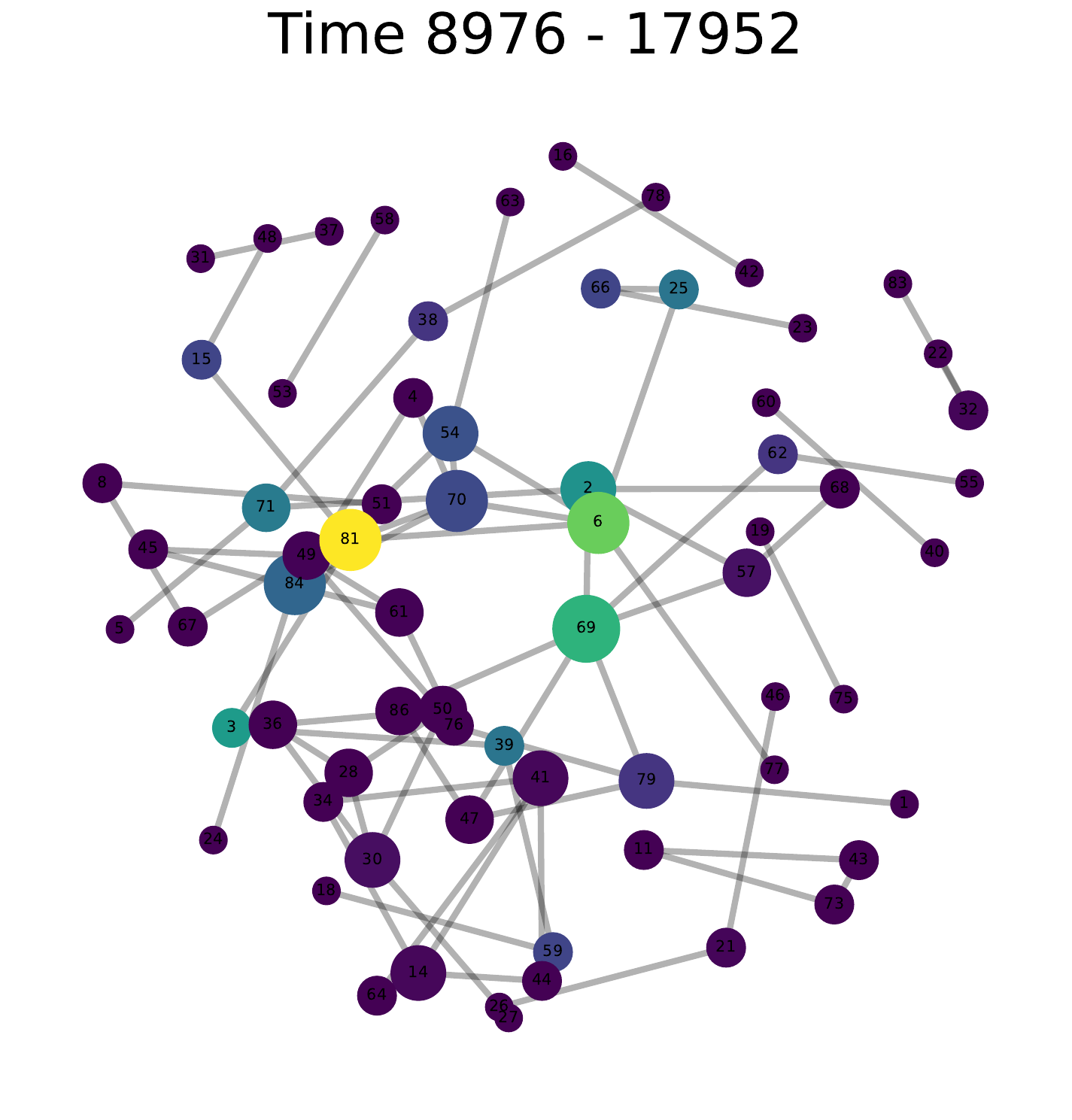}
    \end{minipage}
    \hfill
    \begin{minipage}{0.18\textwidth}
        \centering
        \includegraphics[width=\textwidth]{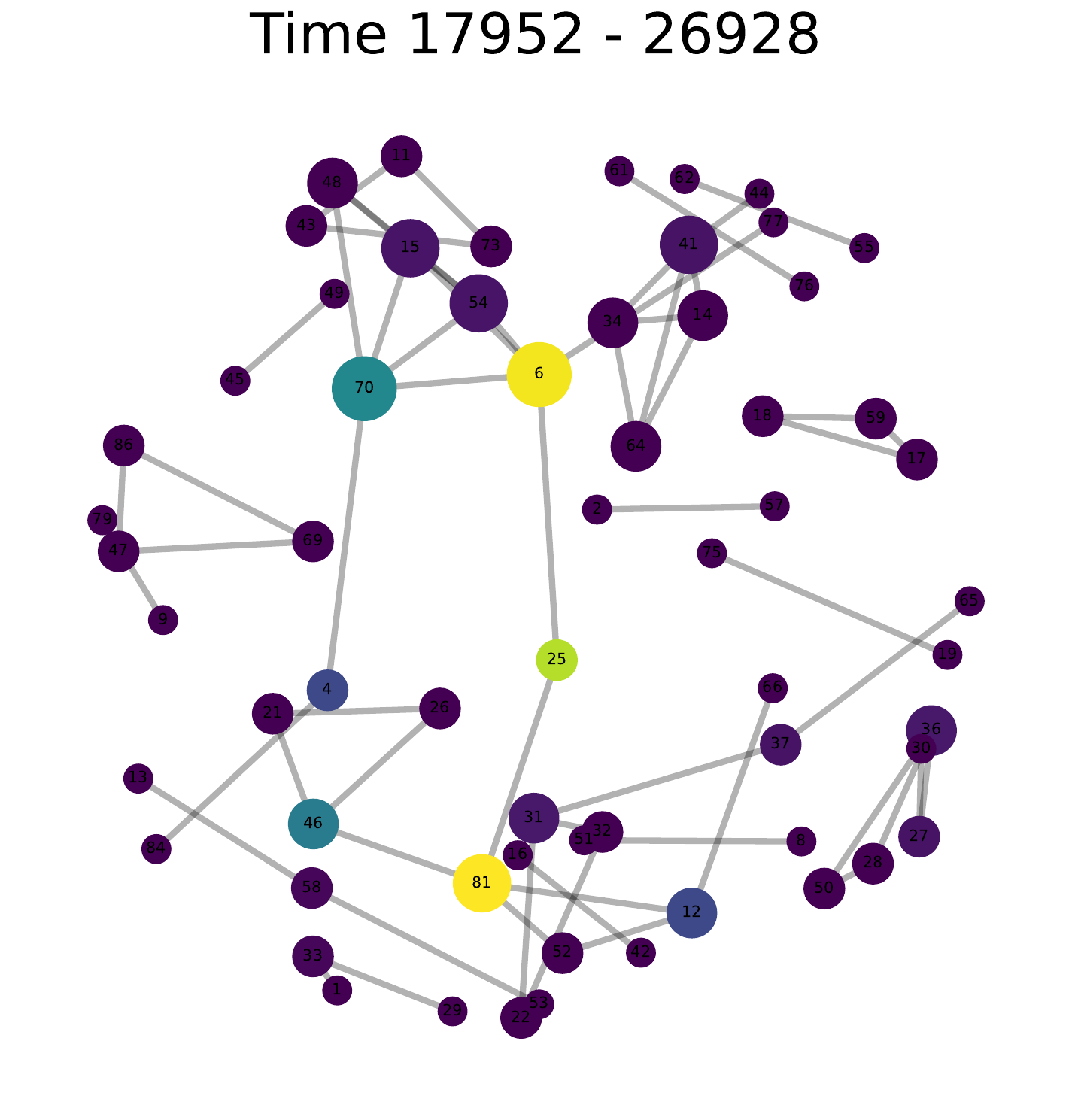}
    \end{minipage}
    \hfill
    \begin{minipage}{0.18\textwidth}
        \centering
        \includegraphics[width=\textwidth]{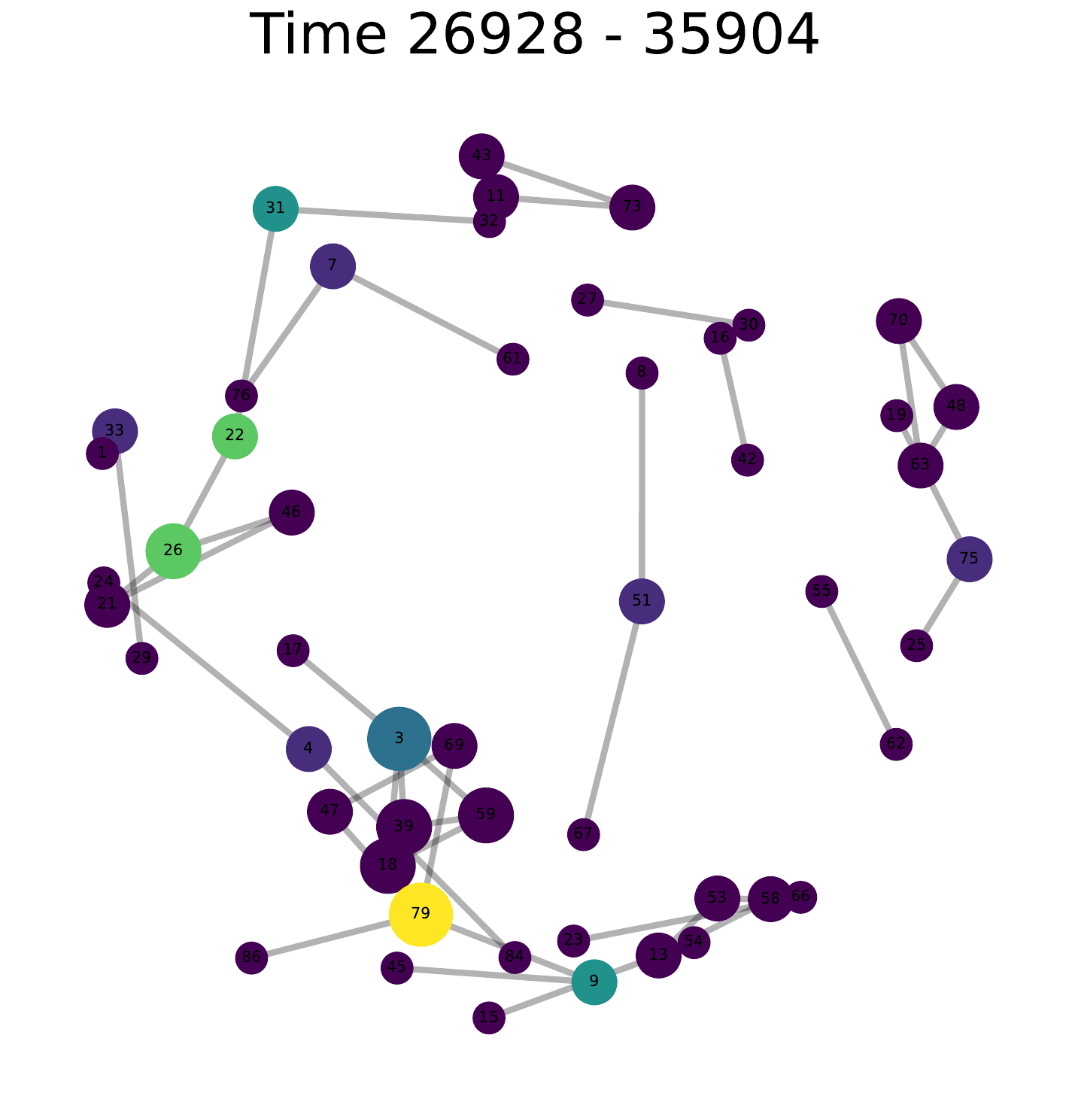}
    \end{minipage}
    \hfill
    \begin{minipage}{0.18\textwidth}
        \centering
        \includegraphics[width=\textwidth]{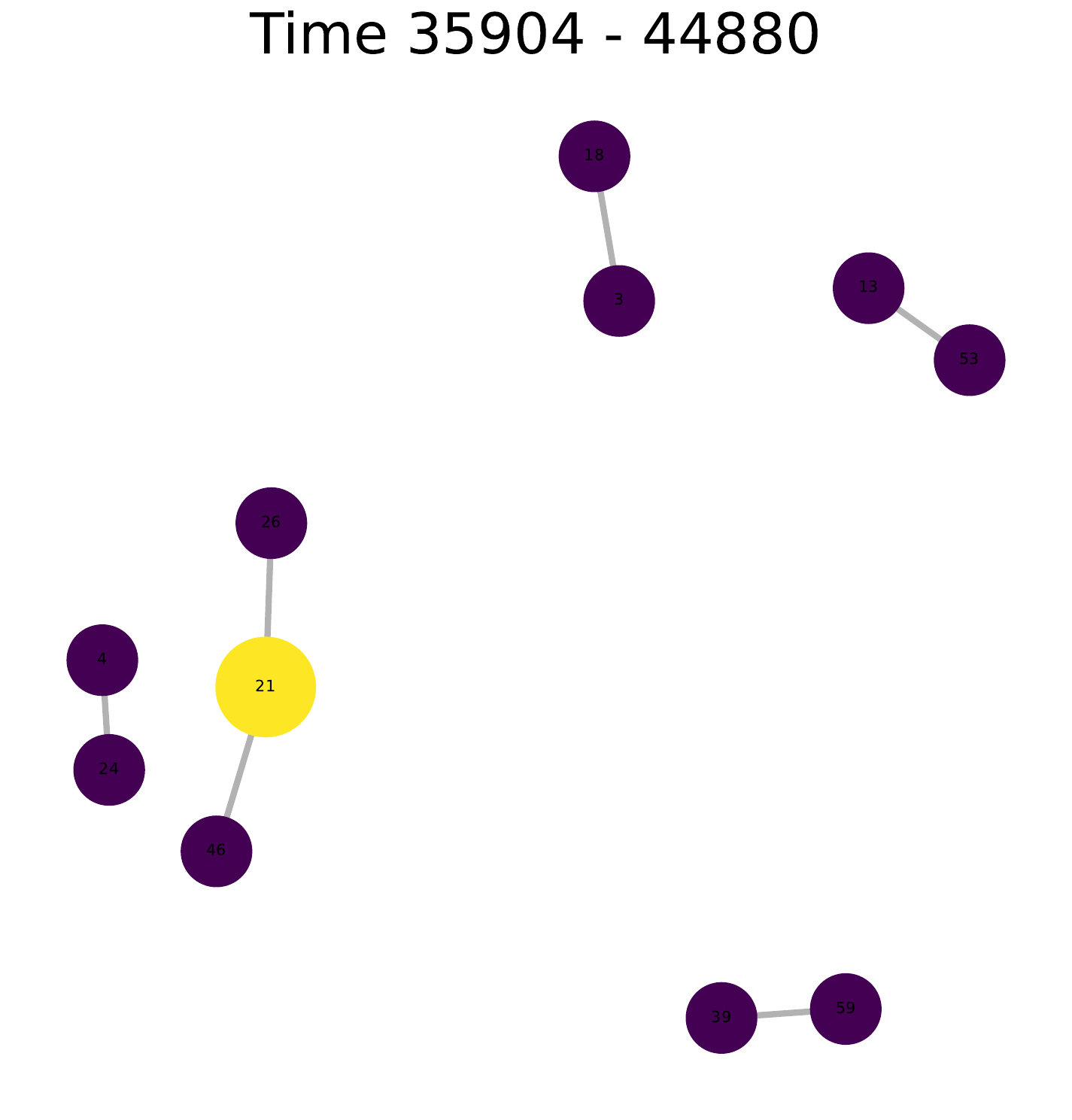}
    \end{minipage}
    
    \caption{Dynamic network evolving with time}
    \label{fig:network_evolving}
\end{figure*}

\begin{figure*}[t]
    \centering
    \includegraphics[width=0.6\textwidth]{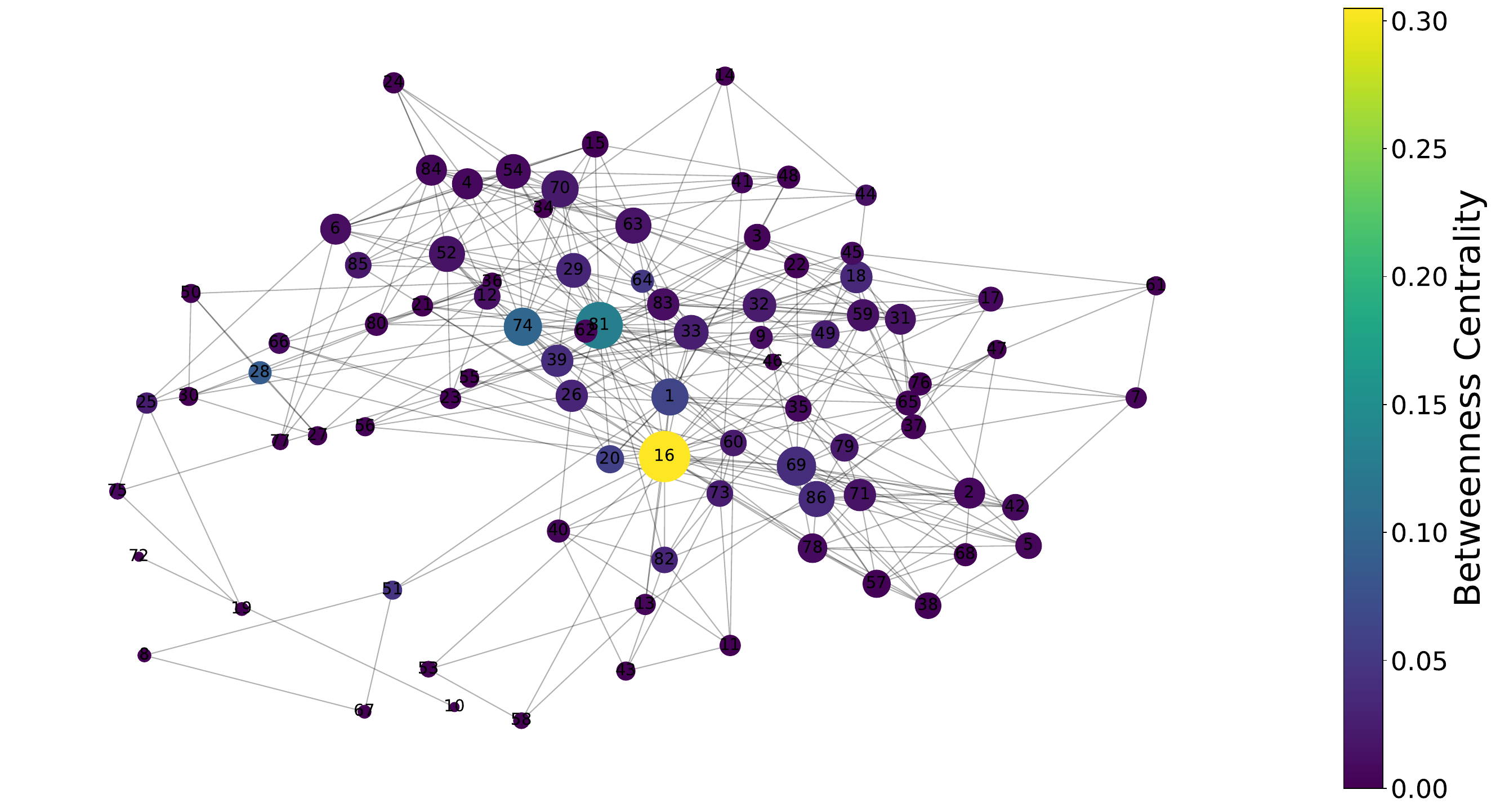}  
    \caption{The characteristics of the nodes in the contact network and the undirected graph.}  
    \label{fig:undirected_graph}  
\end{figure*}

Since static complex networks do not consider the temporal changes in a network, their application is significantly limited. This naturally necessitates the introduction of the concept of dynamic complex networks \cite{lin2020modeling,van2022dynamic,chen2022statistical,wu2018general}. The main difference between dynamic and static complex networks lies in the time dependency, which allows dynamic networks to better reflect the evolutionary characteristics and dynamic behavior of real-world systems.

In a dynamic network, nodes and edges can increase or decrease over time. Mathematically, this can be represented as a series of graphs \( G(t) = (V(t), E(t)) \), where \( t \) denotes time, and \( V(t) \) and \( E(t) \) represent the sets of nodes and edges at time \( t \).
In Figure(\ref{fig:network_evolving}), we depict the network structure of a contact network with 86 nodes over five consecutive time periods in one day \cite{ozella2021using}. From the topological perspective, the nodes and edges of the network undergo significant changes in each time period.
Furthermore, in Figure(\ref{fig:undirected graph}), we construct an undirected graph, where we illustrate the overall network's degree, centrality, betweenness centrality, and closeness centrality. In the figure, the size of the nodes is adjusted according to degree centrality—the larger the node, the more contact it has with other nodes. Node color is determined by betweenness centrality, reflecting the node's "bridging" role in the network. Closeness centrality is used to adjust the distance between a node and other nodes in the network, indicating the node's proximity within the overall network.

Nodes in a dynamic network may adjust their behavior based on the overall state of the network or local environmental changes, leading to feedback and adaptive behavior. For example, species in an ecosystem may adjust their reproductive strategies based on the availability of resources. This adaptability can be modeled by dynamic equations that include the network state, such as \( \mathbf{x}(t+1) = f(\mathbf{x}(t), G(t)) \).
Dynamic networks may exhibit periodic, random, or chaotic temporal evolution patterns. These patterns reflect the complex dynamics of the network over long time sequences. For instance, periodic fluctuations in economic networks can be studied by introducing time series analysis, such as using an autoregressive model \( \mathbf{x}(t) = \phi_1 \mathbf{x}(t-1) + \phi_2 \mathbf{x}(t-2) + \cdots + \epsilon(t) \).
The analysis of dynamic complex networks typically involves a detailed examination of both changes in network topology and the evolution of network dynamics. The topology involves the addition, removal, or alteration of nodes and edges over time. These structural changes can be described and analyzed using various mathematical tools, such as time-dependent graph representations, which can be expressed as a collection of graphs \( \{G(t)\} \), where each graph \( G(t) = (V(t), E(t)) \) represents the set of nodes \( V(t) \) and edges \( E(t) \) at time \( t \).
If the network updates at discrete time points, a series of adjacency matrices \( \{A(t)\} \) can be used to describe it, where \( A(t)_{ij} = 1 \) indicates that there is a connection between nodes \( i \) and \( j \) at time \( t \). For networks that change in real-time or continuously, a time function \( A(t)_{ij} \) can be introduced to represent the existence or strength of the connection between nodes.

The dynamics focus more on how the state of nodes or the entire network evolves over time. This can be described using state equations. For a discrete-time model, a state update equation can be used, such as \( \mathbf{x}(t+1) = f(\mathbf{x}(t), \mathbf{x}(t-1), \dots, G(t)) \), where \( f \) is a function that describes how the system's state evolves based on the current and past states as well as changes in the network structure. For a continuous-time model, differential equations or difference equations can be used, such as \( \frac{d\mathbf{x}(t)}{dt} = g(\mathbf{x}(t), G(t)) \), where \( g \) is a function describing the continuous change of the state over time.

Meanwhile, dynamic and static network models are not completely opposed to each other. When solving practical problems, we can combine the two to improve the overall performance of the model. For example, embedding static structural features in dynamic networks \cite{jin2022generalizing} or introducing dynamic key points into static networks \cite{yu2020identifying} to enhance efficiency.
The issue explored in this article is how to capture the characteristics of information propagation in dynamic complex networks. This can be done by constructing a set of mathematical models to describe how information is transmitted within a network structure that changes over time.

Firstly, the network model is defined similarly to the previous definitions:
A dynamic graph can be represented as \( G(t) = (V(t), E(t)) \), where \( V(t) \) is the set of nodes at time \( t \), \( E(t) \) is the set of edges at time \( t \), and \( A(t) \) is the corresponding adjacency matrix, where \( A(t)_{ij} = 1 \) indicates a direct connection between nodes \( i \) and \( j \) at time \( t \).
The process of information propagation can be described by the state of each node \( i \) at time \( t \), denoted as \( x_i(t) \). The state \( x_i(t) \) can take various forms, such as a binary state indicating the presence or absence of information (with 1 representing the presence of information and 0 representing its absence) or more complex forms, such as the strength or confidence of the information.

The propagation of information can be defined by a discrete-time dynamic update equation:
\begin{equation}
    x_i(t+1) = f(x_i(t), \{x_j(t) | j \in N_i(t)\}, t)
\end{equation}

\noindent where \( N_i(t) \) is the set of neighbors of node \( i \) at time \( t \), and \( f \) is a propagation function that describes how the information state of node \( i \) is updated at the next time step based on its current state and the states of its neighbors.

The function \( f \) can be further refined to capture specific mechanisms of information propagation, such as probabilistic spreading:

\begin{equation}
    x_i(t+1) = \begin{cases} 
    1 & \text{if } \sum_{j \in N_i(t)} w_{ij}(t) x_j(t) + b_i(t) > \theta_i(t) \\
    x_i(t) & \text{otherwise}
    \end{cases}
\end{equation}

\noindent where \( w_{ij}(t) \) represents the influence weight of node \( j \) on node \( i \) at time \( t \),
\( b_i(t) \) is the influence of external information sources on node \( i \),
\( \theta_i(t) \) is the threshold that determines whether node \( i \) updates its information state.

Finally, the dynamic adjustment of parameters is considered. In some cases, the propagation weight \( w_{ij}(t) \), the threshold \( \theta_i(t) \), and other parameters may change dynamically over time and with the network state to simulate the network’s and nodes' adaptability to environmental changes:

\begin{equation}
    w_{ij}(t+1) = g(w_{ij}(t), x_i(t), x_j(t), t)
\end{equation}

Through such mathematical definitions, researchers can precisely simulate and analyze the behavior of information propagation in dynamic complex networks, understanding how various factors influence the speed and reach of information diffusion. How to capture network characteristics in specific network problems based on this framework is the fundamental question addressed in this research.

\section{Model Based Method}
\begin{figure*}[t]
    \centering
    \begin{subfigure}{0.45\textwidth}
        \centering
        \includegraphics[width=\textwidth]{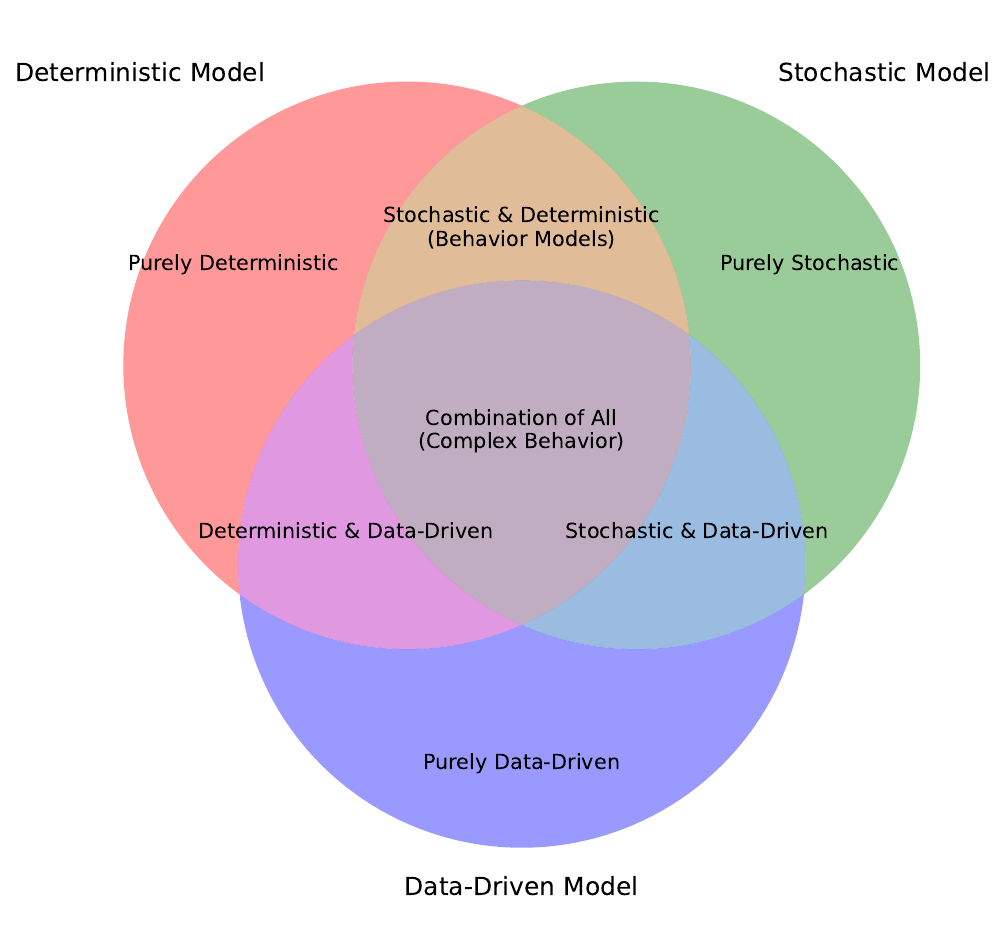}  
        \label{fig:subfig1}  
    \end{subfigure}
    \hfill
    \begin{subfigure}{0.45\textwidth}
        \centering
        \includegraphics[width=\textwidth]{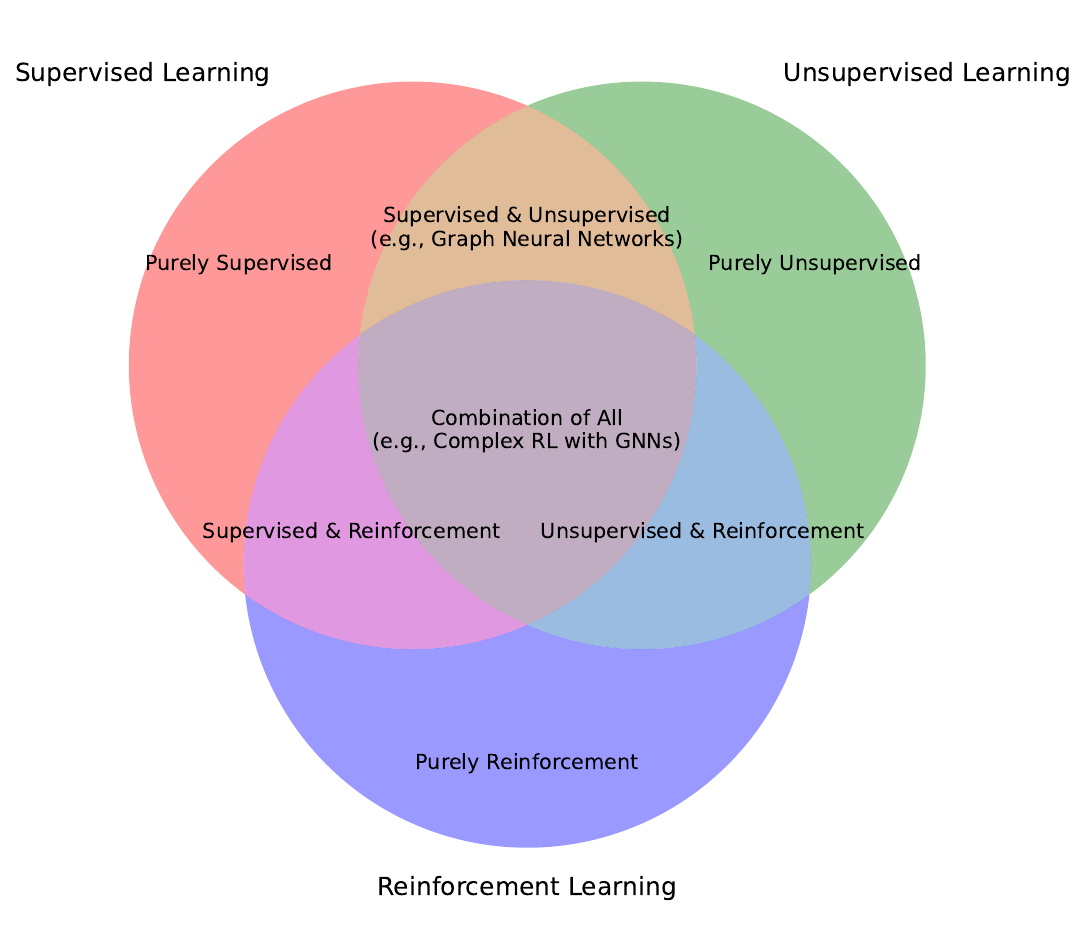}  
        \label{fig:subfig2}  
    \end{subfigure}
    \caption{Relationship and Overlap Between Different Model-based and Model-free Approaches}  
    \label{fig:overlap_relation}  
\end{figure*}

The methods discussed in this paper for addressing complex network propagation issues can be broadly divided into two categories: model-based methods and model-free methods. We show the overlap relationship of different approaches in figure(\ref{fig:overlap_relation}). The techniques and theoretical foundations they use differ significantly.
Model-based methods rely on an explicit model of the system or process. This model can be physical, statistical, mathematical, or computational and is used to describe the behavior or dynamics of the network system. These models are typically based on an understanding of the system's operating mechanisms, including its structure, function, and interactions.
The advantage is that they provide deep insights into the system, allowing for predictions of its behavior under different conditions, as well as enabling optimization and control design since the model offers insights into the system's responses. 
However, modeling can be complex and time-consuming, especially when the system is highly complex or not well understood. The accuracy of the model also depends on the initial assumptions and the quality of the available data.

On the other hand, "model-free" methods do not rely on explicit modeling of the system. These methods learn or control the network system directly from input-output data without attempting to understand the internal workings of the system.
The advantages of this approach include high flexibility, making it suitable for systems where modeling is difficult or unknown. It can learn directly from real-world data and is well-suited for handling highly complex and nonlinear problems. However, the lack of understanding of the system's internal mechanisms may lead to poor interpretability of the solutions, and a large amount of data may be required to achieve effective learning.
This section will discuss some model-based methods in detail, while model-free methods will be covered in the next chapter. The comparison of the methods discussed in this section and the paper can be found in Table \ref{table: model_based}.

\subsection{Deterministic Model}
A deterministic model is a type of model that produces the same results each time it is run under the same initial conditions. This means that the system's behavior is entirely determined by the initial conditions and the parameters of the model, with no randomness or uncertainty involved \cite{piccirillo2021nonlinear,moore2020predicting,zarin2022deterministic,mei2017dynamics}. This is also a primary paradigm used by researchers in the early stages of addressing network problems \cite{kermack1927contribution,sarkar2003survey}.
Using a deterministic model to capture the dynamics of propagation in a network is an important method for studying the spread of information, viruses \cite{mei2017dynamics,zarin2022deterministic}, rumors \cite{chen2022information}, and other behaviors in complex networks \cite{de2021reduced}. Deterministic models typically describe the changes in state and dynamic behavior during the propagation process through a set of differential or difference equations.Let's further explain how to use a deterministic model to model the propagation process in a network using the SIR model as an example.

A network can be represented by a graph \(G(V, E)\), where \(V\) is the set of nodes representing individuals or entities in the system, and \(E\) is the set of edges representing the connections or interactions between these entities.
In the SIR model \cite{kermack1927contribution}, each node represents an individual and can be in one of the following three states:
Susceptible (S), a node that can potentially be infected.
Infected (I), a node that is already infected and capable of spreading the disease.
Recovered (R), a node that has recovered, no longer spreads the disease, and cannot be reinfected.
At the start, most nodes in the network are usually initialized as susceptible (S), while a few nodes are set as initially infected (I).
The propagation process is controlled by two key parameters:
Transmission rate (\(\beta\)), the probability that an infected individual will spread the disease or information to a susceptible individual. It controls the rate at which susceptible nodes become infected.
Recovery rate (\(\gamma\)), the rate at which infected individuals recover and move into the recovered state.
On the network, the transmission rate may be related to the network structure. For example, if node \(i\) is connected to multiple infected individuals, its probability of infection can be seen as the cumulative effect of independent transmission events from each of these connections.
At each time step, the propagation process proceeds according to the following rules:
1. Infection Spread: For each susceptible node \(i\), the probability of being infected is related to whether any of its neighboring nodes \(j\) are infected and the transmission rate \(\beta\). The probability that a susceptible node \(i\) becomes infected can be expressed as:

\begin{equation}
    P(\text{Susceptible node } i \text{ being infected}) = 1 - \prod_{j \in N(i)} (1 - \beta A_{ij} I_j)
\end{equation}
   
\noindent where \(N(i)\) represents the set of neighbors of node \(i\), \(A_{ij}\) is the element of the adjacency matrix indicating the connection between nodes \(i\) and \(j\), and \(I_j\) is the infection status of node \(j\) (1 if \(j\) is infected, otherwise 0).
Recovery, Each infected individual transitions from the infected state to the recovered state at a rate governed by the recovery rate \(\gamma\).

At each time step, the states of all nodes are updated according to the propagation rules.
Susceptible nodes may become infected, Infected nodes may recover and become recovered.
This process continues until no new infections occur in the network, or a predefined number of time steps is reached.

\begin{sidewaystable}[htbp]
\centering
\begin{tabularx}{\textwidth}{|m{3cm}|m{4cm}|X|X|X|}
\hline
\textbf{Method Type} & \textbf{References} & \textbf{Advantages} & \textbf{Disadvantages} & \textbf{Application Areas} \\ \hline
\centering Deterministic Model & \centering  \cite{piccirillo2021nonlinear,moore2020predicting,zarin2022deterministic, mei2017dynamics, kermack1927contribution, sarkar2003survey, chen2022information} & Predictable, no randomness, provides clear insight into system dynamics & Simplistic, may not capture real-world uncertainties & Epidemiology, Network Propagation, Information Spread \\ \hline
\centering Stochastic Model & \centering  \cite{shah2024stochastic, lu2022analytical, scussel2021stochastic, olabode2021deterministic, ren2018stochastic, zhu2018stochastic, jain2019stochastic, alem2016stochastic, long2019dynamic, correia2023stochastic, jin2022stochastic} & Accounts for uncertainty, handles real-world randomness effectively & Less predictable, harder to analyze, requires more computational resources & Communication Networks, Social Networks, Transportation Networks \\ \hline
\centering Behavior-Based Model & \centering  \cite{de2023framework, subramanian2024socio, shan2024diffusion, von2023framework, hashimoto2022agent, huang2022overview, deng2021user} & Captures complex individual behaviors, simulates interactions & Complex, highly dependent on assumptions about behaviors and interactions & Agent-based modeling, Opinion Dynamics, Decision-making systems \\ \hline
\centering Data-Driven Model & \centering  \cite{liu2022survey, chen2022recent, an2015practical, xue2020data, zhu2018big, fang2020transmission, pierri2019false} & Learns directly from data, highly flexible and can adapt to real scenarios & Requires large datasets, prone to overfitting, sensitive to data quality & Big Data Analytics, Social Network Analysis, Epidemic Forecasting \\ \hline
\end{tabularx}
\caption{Comparison of Model-based Methods for Network Propagation}
\label{table: model_based}
\end{sidewaystable}

By simulating the above propagation process multiple times, one can analyze characteristics of the spread, such as:
Basic Reproduction Number (\(R_0\)), which reflects the average number of other susceptible individuals that one infected individual can infect initially.
Spread Rate, the rate of change in the number of infected individuals over time.
Final Infection Ratio, the proportion of nodes that have been infected by the end of the propagation.
Impact of Network Structure on Propagation, for instance, densely connected groups (clusters) may have a higher infection rate.
Since the spread in the SIR model on a network is often difficult to solve analytically, numerical simulation is a commonly used method to study propagation dynamics. The propagation process can be simulated using discrete time steps, with the states of all nodes updated at each time step.
\subsection{Stochastic Model}
A stochastic model is a type of model that takes into account uncertainty or randomness in the system \cite{shah2024stochastic, lu2022analytical, scussel2021stochastic}. This model assumes that the system's future state depends not only on the current state and known parameters but also on random factors, so even with the same initial conditions, different results may be produced each time the model is run.
Stochastic models and deterministic models are two opposing concepts, both of which are widely used in addressing network problems \cite{olabode2021deterministic,chen2022information}.
The introduction of the concept of random variables into deterministic models can avoid many of the limitations of deterministic models, leading to widespread applications in fields such as communication networks \cite{ren2018stochastic,zhu2018stochastic}, social networks \cite{jain2019stochastic}, transportation networks \cite{alem2016stochastic, long2019dynamic}, and more \cite{correia2023stochastic,jin2022stochastic}.
Let's take random walks \cite{kogias2009study} and diffusion processes as examples.

The network definition is similar to what was previously described.
In a random walk model, each node can be considered as a possible location for a walker, and at each time step, the walker randomly selects any one of the nodes connected to the current node and moves there.

Let \( X_t \) represent the node where the walker is located at time \( t \). The transition probability of \( X_t \) is given by the matrix \( P \), where \( P_{ij} \) is the probability of moving from node \( i \) to node \( j \). If the network is undirected and all edges are equally important, \( P_{ij} \) is defined as:
\begin{equation}
    P_{ij} = \begin{cases} 
    \frac{1}{d_i} & \text{if } A_{ij} = 1 \\
    0 & \text{otherwise}
    \end{cases}
\end{equation}

Here, \( d_i \) is the degree of node \( i \), i.e., the number of nodes directly connected to node \( i \).

The diffusion process is a natural extension of the random walk model, used to simulate how a certain substance or information spreads through a network starting from one or more source nodes.
Let \( p_i(t) \) represent the probability that node \( i \) contains or is affected by the substance at time \( t \). Initially, one or more source nodes are selected, and the \( p_i(0) \) for these nodes is set to 1, while the rest are set to 0. Over time, this probability spreads through the network, and the update rule can be expressed as:

\begin{equation}
    p_i(t+1) = \sum_{j \in N(i)} P_{ji} p_j(t)
\end{equation}

\noindent Here, \( N(i) \) is the set of neighbors of node \( i \), and \( P_{ji} \) is the transition probability from node \( j \) to node \( i \).

In some cases, we may be interested in situations where the walker is "absorbed" (e.g., information is forgotten or a virus is eradicated). This can be simulated by introducing absorbing states and corresponding transition probabilities.
Long-term behavior analysis often involves calculating the stationary distribution, that is:

\begin{equation}
    \lim_{t \to \infty} p_i(t)
\end{equation}

This reflects the probability of a certain piece of information or state existing at each node after a long period of time.
Through computer simulations, we can track the actual spread path and speed of information or viruses in the network, thereby analyzing how the network structure affects the spread. This model is particularly useful for studying phenomena such as information diffusion in social networks or disease spread in populations.

\subsection{Behavior Based Model}

A behavior-based model is a method for simulating and analyzing the behavior patterns of individuals or groups. 
This method is more like a way of thinking about solving network problems rather than a method \cite{stolfo2006behavior,hashimoto2022agent,gong2021transaction}.
It emphasizes understanding the overall behavior of a system by focusing on individual behaviors \cite{de2023framework,subramanian2024socio,shan2024diffusion}, and simulating various decisions and behavioral responses.

Let's take agent-based models (ABM) as an example \cite{von2023framework,hashimoto2022agent}. This type of model allows each agent (i.e., individual in the model) to have its own attributes and behavioral rules, enabling them to interact within a simulated environment.
In ABM, nodes represent agents, and edges represent the connections between nodes (such as social ties, communication channels, etc.). Each node \( i \in V \) can have a range of attributes, such as age, gender, social status, and more.
Each agent \( i \) has a set of attributes \( A_i \), which influence the agent's behavior and decision-making. These attributes can include state information (such as susceptible, infected, or recovered in a disease transmission model), opinions (in opinion dynamics models), or any other relevant characteristics.
A function \( B_i(t) \) is defined to describe the behavior of each agent \( i \) at time \( t \). These behaviors may depend on the agent's internal state, the state of its neighbors, or global environmental factors. Behaviors could include:
State updates, for instance, in an infectious disease model, an agent may change its health status based on the proportion of infected neighbors.
Decision-making, Agents might make decisions based on specific thresholds or probabilities (e.g., adopting new technology or changing opinions).
Behavioral rules can be expressed using a probability \( P_{i \to j} \), representing the probability that agent \( i \) influences or is influenced by its neighbor \( j \).
Interactions between agents are typically based on network connections. Let \( N(i) \) denote the set of neighbors of node \( i \), and the interaction process can be described as:
\begin{equation}
    S_i(t+1) = f(S_i(t), \{S_j(t) | j \in N(i)\}, \epsilon_i(t))
\end{equation}

\noindent where \( S_i(t) \) is the state of agent \( i \) at time \( t \), and \( f \) is a function describing how the agent's state is updated based on the states of its neighbors and possible random events \( \epsilon_i(t) \).
The dynamic evolution of the model is typically achieved through iterative time steps, where at each step, all agents' states are updated according to the behavioral and interaction rules. This iteration can be either synchronous (all agents update simultaneously) or asynchronous (agents update one by one).
By altering parameters (such as connection probability, thresholds, etc.) and observing their impact on the model's behavior, one can perform parameter tuning and sensitivity analysis to better understand the key drivers of network propagation. Based on this, it has various applications. \cite{hashimoto2022agent,huang2022overview,deng2021user}

\subsection{Data-Driven Models}

A data-driven model does not rely on traditional theoretical frameworks to establish its structure but instead learns and infers model parameters directly from large amounts of real-world data. It can be either a model-based or model-free approach \cite{liu2022survey,chen2022recent,an2015practical}.
Model-based methods typically depend on some form of a predefined model that describes the relationship between inputs and outputs. These models may be derived from theoretical principles, empirical rules, or obtained through data fitting.
In such methods, the parameters of the model need to be estimated based on the available data. 
This approach leverages actual data to refine and validate models \cite{xue2020data,zhu2018big,fang2020transmission,pierri2019false}, ensuring that they more accurately reflect the underlying processes or phenomena being studied.
Let's interpret this approach by combining it with the SIR model discussed earlier \cite{barmparis2020estimating,kermack1927contribution}.
The dynamics of the SIR model can be described by the following differential equations:

\begin{subequations}
    \begin{equation}
    \frac{dS}{dt} = -\beta(t) S I
    \end{equation}
    \begin{equation}
    \frac{dI}{dt} = \beta(t) S I - \gamma(t) I
    \end{equation}
    \begin{equation}
    \frac{dR}{dt} = \gamma(t) I
    \end{equation}
\end{subequations}

\noindent Here, the transmission rate \( \beta(t) \) and recovery rate \( \gamma(t) \) are functions learned from data, potentially depending on time and other network characteristics.
For each node, we can extract its network attributes, such as degree, betweenness centrality, clustering coefficient, etc., or global features of the social network, such as network density, average path length, and so on. We then collect data on \( S \), \( I \), and \( R \) at different time points as the training targets.
By using historical data from past propagation events, the model can be trained to predict \( \beta(t) \) and \( \gamma(t) \). Methods such as cross-validation can be employed to verify the predictive performance of the model.
In a simulated environment, the learned \( \beta(t) \) and \( \gamma(t) \) can be used to dynamically simulate the process of information spread. This allows us to analyze how different network structures and parameter settings influence the effectiveness of information propagation, helping to optimize dissemination strategies or predict future trends.


\section{Model-Free Method}
In this section, we will focus on using model-free methods to capture the characteristics of network propagation. The definition of model-free methods varies in different contexts. Here, we are focusing on methods of machine learning based on neural networks \cite{abiodun2018state,jordan2015machine,wu2024state} or deep learning \cite{dong2021survey,pouyanfar2018survey}. More specifically, we will explore the application of supervised learning \cite{muhammad2015supervised,jiang2020supervised}, unsupervised learning \cite{dike2018unsupervised,khanum2015survey,schmarje2021survey}, reinforcement learning \cite{arulkumaran2017deep,shakya2023reinforcement}, and graph neural networks \cite{wu2020comprehensive,zheng2022graph} on network propagation. The comparison of the methods discussed in this section and the paper can be found in Table \ref{table:model-free based}.
\subsection{Supervised Learning}

\begin{sidewaystable}[htbp]
\centering
\rotatebox{0}{ 
\begin{minipage}{\textheight} 
\centering
\begin{tabularx}{\textwidth}{|m{3cm}|m{4cm}|X|X|X|}
\hline
\textbf{Method Type} & \textbf{References} & \textbf{Advantages} & \textbf{Disadvantages} & \textbf{Application Areas} \\ \hline
\centering Supervised Learning & \centering  \cite{mladenovic2022overview,zhang2020cellular,panayiotou2016performance,he2020random,chen2019lstm,gou2022message,datilo2019review,ye2020methodology} & Learns from labeled data, allows for precise predictions & Requires large amounts of labeled data, may not generalize well to unseen data & Social Networks, Epidemic Forecasting, Network Propagation \\ \hline
\centering Unsupervised Learning & \centering  \cite{wang2022predicting, qourbani2023toward, zideh2024unsupervised, azcorra2018unsupervised, choi2019unsupervised} & Can discover hidden patterns without labeled data & Harder to evaluate results, may require more data to capture significant structures & Data Clustering, Pattern Recognition, Network Structure Analysis \\ \hline
\centering Reinforcement Learning & \centering  \cite{ghavipour2018dynamic,wang2021deep,wu2019decentralized,libin2021deep,he2022reinforcement,lee2019reinforcement} & Learns through trial and error, good for decision-making in dynamic environments & Requires careful reward design, may need significant computational resources & Social Networks, Multi-Robot Systems, Internet of Things \\ \hline
\centering Graph Neural Networks & \centering  \cite{yu2021self,cao2022mepognn,mahmud2021human,sivakumar2023enhancing,dong2023graph,li2023graph} & Effectively captures complex node relationships, excels at structured data tasks & Computationally expensive, requires large amounts of data to generalize & Social Networks, Complex Networks, Network Propagation \\
\bottomrule
    \end{tabularx}
    \caption{Comparison of Model-free Methods for Network Propagation}
    \label{table:model-free based}
\end{minipage}
}
\end{sidewaystable}

In supervised learning, the model learns how to map inputs to outputs by analyzing the relationship between the input data (features) and the output labels (targets) \cite{mladenovic2022overview,zhang2020cellular,panayiotou2016performance}.
To capture the characteristics of information dissemination in a network using supervised learning, we need to clearly define what specific aspects of information dissemination are of interest \cite{he2020random,chen2019lstm}. For example, we might be interested in predicting the speed of information spread in the network \cite{gou2022message}, the scope of dissemination (the number of nodes affected) \cite{datilo2019review}, or the types of nodes that will eventually be influenced \cite{ye2020methodology}.

Suppose our goal is to predict the number of nodes that a message can reach within a specific time \( T \) after it starts spreading from the source node.
Without using graph neural networks (GNNs), one could consider representing the network's structural information and node characteristics in a non-graphical manner:
For example, the adjacency matrix representation, where the network structure is represented by the adjacency matrix \( A \), with \( A_{ij} \) indicating the connection status between node \( i \) and node \( j \) (e.g., 1 for connected, 0 for not connected).
Alternatively, the feature matrix representation, where each node's attributes can be represented by a feature matrix \( X \), with each row representing the feature vector of a node.
A deep learning model, such as a multilayer perceptron (MLP), can be used to process structured data. The model's input could be the feature vector of a node, combined with an aggregation of the feature vectors of its neighboring nodes (e.g., through weighted averaging).

Let \( x_i \) be the feature vector of node \( i \), then the input representation of node \( i \) can be defined as:

\begin{equation}
     v_i = f\left( x_i, \sum_{j \in \mathcal{N}(i)} w_{ij} \cdot x_j \right) 
\end{equation}

\noindent where \( f \) is a nonlinear function (such as ReLU), \( \mathcal{N}(i) \) denotes the set of neighboring nodes of node \( i \), and \( w_{ij} \) is the weight from node \( j \) to \( i \) (which may be pre-set or learned).

The objective function (loss function) of the model can be defined as:

\begin{equation}
    \mathcal{L}(\theta) = \sum_{i=1}^N \left( y_i - \hat{y}_i(\theta) \right)^2
\end{equation}

\noindent where \( \theta \) represents the model parameters, \( y_i \) is the actual label (e.g., the number of nodes affected within time \( T \)), and \( \hat{y}_i(\theta) \) is the model's predicted output.
The model is trained by minimizing the loss function on the training set, and then validated and tested on the test set to evaluate the model's generalization ability on new data.
This approach is a simple paradigm for solving network dissemination problems, such as those in social networks or propagation networks.
\subsection{Unsupervised Learning}
Unsupervised learning focuses on discovering hidden structures and patterns from unlabeled data \cite{wang2022predicting, qourbani2023toward}. Unlike supervised learning, the dataset in unsupervised learning does not provide output labels or results, so the model must independently explore the intrinsic properties and relationships within the data \cite{zideh2024unsupervised, qourbani2023toward,chen2018unsupervised,azcorra2018unsupervised,choi2019unsupervised}. 
In many scenarios, considering algorithm efficiency, introducing unsupervised learning on the basis of supervised learning can effectively improve the overall efficiency of the model \cite{wang2022predicting, gupta2023multi}.

When utilizing unsupervised learning based on deep learning \cite{hwang2020unsupervised,gangireddy2020unsupervised} to capture the characteristics of information propagation in a network, Generative Adversarial Networks (GANs) offer an intriguing approach \cite{wu2021ep,wang2022predicting}. GANs work by training two networks—one generator and one discriminator—that compete against each other to generate new data that resembles real data. In the context of network information propagation, we can imagine using GANs to generate simulated propagation patterns to understand the dynamics of propagation under different conditions.

In this setup, our goal is to train a model that generates simulated samples of information propagation in the network, which should be statistically indistinguishable from real propagation patterns. For instance, we may be interested in how information spreads across a social network, including the number of affected nodes, the propagation paths, and more.
First, we need to define the representation of the network and the representation of information propagation:
We can describe the structure of the network and the characteristics of the nodes using an adjacency matrix \( A \) or a feature matrix \( X \).
At the same time, we define a vector \( p \), where \( p_i \) represents the presence or activity level of information at node \( i \).

The goal of the generator \( G \) is to generate fake information propagation patterns \( \tilde{p} \) based on input noise \( z \) (usually a random vector):

\begin{equation}
    \tilde{p} = G(z; \theta_g) 
\end{equation}

\noindent where \( \theta_g \) represents the parameters of the generator.

The goal of the discriminator \( D \) is to distinguish whether the input information propagation pattern comes from real data \( p \) or is generated by the generator \( \tilde{p} \):

\begin{equation}
    D(p; \theta_d)
\end{equation}

\noindent where \( D(p; \theta_d) \) outputs a probability value indicating the likelihood that \( p \) is a real propagation pattern. \( \theta_d \) represents the parameters of the discriminator.
Training the GAN involves minimizing and maximizing an objective function, known as the adversarial loss:
\begin{equation}
    \min_{\theta_g} \max_{\theta_d} \mathbb{E}_{p \sim \text{data}}[\log D(p)] + \mathbb{E}_{z \sim \text{noise}}[\log(1 - D(G(z)))]
\end{equation}

\noindent Here, the generator attempts to produce propagation patterns as realistic as possible to fool the discriminator, while the discriminator strives to accurately distinguish between real and generated data.
Through training, GANs can learn the complex distribution within the data, which may include characteristics such as propagation paths, speed, and impact of propagation. Once trained, the generator can be used to generate new information propagation patterns for further analysis and research.
This GAN-based approach allows us to delve into understanding and simulating the process of information propagation in complex networks without explicit labels. By altering the network structure or the initial conditions of information propagation, we can also explore the impact of different factors on the outcome of information propagation.


\subsection{Reinforcement Learning}
Reinforcement learning is a unique machine learning paradigm that enables an agent to learn behavior strategies in an environment through trial and error, with the goal of maximizing its cumulative reward \cite{ghavipour2018dynamic,wang2021deep}. Unlike supervised and unsupervised learning, reinforcement learning specifically focuses on how to make decisions in the absence of explicit instructions on the correct actions to take.

When utilizing deep reinforcement learning (DRL) \cite{libin2021deep,fan2020finding,wang2021deep} to capture the characteristics of information propagation in a network, we can build a model where an agent learns how to effectively propagate information through exploration and exploitation of strategies. 
This method is not only widely applied in traditional social networks \cite{he2022reinforcement, wang2020risk} and disease network analysis \cite{libin2021deep}, but is also extensively used in multi-robot systems \cite{wu2019decentralized,wu2024deep}, intelligent transportation networks \cite{lee2019reinforcement,wang2021deep}, and the Internet of Things \cite{wang2020risk}.

This approach can be used to optimize information dissemination strategies, such as maximizing the coverage of information in a social network \cite{he2022reinforcement,wang2020risk} or minimizing transmission costs in a sensor network \cite{lee2019reinforcement}.

First, define the following key components:
Environment: This includes the entire network and its dynamics, such as the state of nodes and their connections.
State (S): The current state of information propagation in the network, which can be represented as a vector of the information reception states of all nodes.
Action (A): The actions that the agent can perform, such as selecting one or more nodes to start the information propagation or pushing a specific message. 
Reward (R): The feedback based on the effect of the action, such as the number of reached nodes, the quality of information reception, or the cost of propagation.

\begin{sidewaystable}[htbp]
\centering
\rotatebox{0}{ 
\begin{minipage}{\textheight} 
    \centering
    \begin{tabularx}{\textwidth}{l c c c c c c c c}
\toprule
\textbf{Method} & \textbf{Complexity} & \textbf{Accuracy} & \textbf{Data Dependence} & \textbf{Robustness} & \textbf{Generalization} & \textbf{Scalability} & \textbf{Stability} & \textbf{Explainability} \\ \midrule
Deterministic & Low & Medium & Low & Low & Low & High & High & High \\ 
Stochastic & Medium & High & Medium & High & Medium & Medium & Medium & Medium \\ 
Behavior-Based & High & Medium & Medium & Medium & Medium & Low & Medium & Medium \\ 
Data-Driven & Medium & High & High & Medium & High & Medium & Medium & Low \\ 
Supervised & Medium & High & High & Medium & Medium & Medium & High & Medium \\ 
Unsupervised & Medium & Medium & High & Medium & High & Medium & Medium & Low \\ 
Reinforcement & High & High & Medium & Medium & High & Medium & Low & Medium \\ 
GNN & High & High & High & Medium & High & Medium & Medium & Low \\ 
\bottomrule
    \end{tabularx}
    \caption{Evaluation of Different Methods for Network Propagation}
    \label{table:evaluation}
\end{minipage}
}
\end{sidewaystable}

Let \( S \) be the state space, where each state \( s \in S \) describes the information reception state of each node in the network. Let \( A \) be the action space, where each action \( a \in A \) describes the nodes chosen by the agent for information propagation.

The reward function \( R(s, a) \) is defined as the immediate reward obtained after taking action \( a \) in state \( s \). For example, the reward can be the number of nodes that newly receive the message minus the cost incurred by the action.
The policy function \( \pi(a|s) \) is a probability distribution that defines the probability of selecting action \( a \) given the state \( s \).
The value function \( V^\pi(s) \) represents the expected cumulative discounted reward starting from state \( s \) and following the policy \( \pi \). It is mathematically defined as:

\begin{equation}
    V^\pi(s) = \mathbb{E}\left[\sum_{t=0}^{\infty} \gamma^t R(s_t, a_t) \mid s_0 = s, \pi\right]
\end{equation}

\noindent where \( \gamma \) is the discount factor, between 0 and 1, which adjusts the current value of future rewards.

Using DRL algorithms, such as Deep Q-Networks (DQN) or policy gradient methods, the agent can learn how to select actions that maximize future cumulative rewards. These algorithms achieve this by interactively exploring the state space and optimizing the value function or policy function.
In DQN, a deep neural network is used to approximate the optimal value function \( Q^*(s, a) \), and the network is trained by minimizing the following loss function:
\begin{equation}
    L(\theta) = \mathbb{E}\left[\left(R(s, a) + \gamma \max_{a'} Q(s', a'; \theta^-) - Q(s, a; \theta)\right)^2\right]
\end{equation}

\noindent where \( \theta \) represents the network parameters, \( \theta^- \) are the parameters of the target network, and \( s' \) is the new state after taking action \( a \).

Through this setup, a DRL-based model can learn how to effectively propagate information within the network, adapt to changes in the environment, and optimize the information dissemination strategy.

\begin{figure*}[t]
    \centering
    \includegraphics[width=0.9\textwidth]{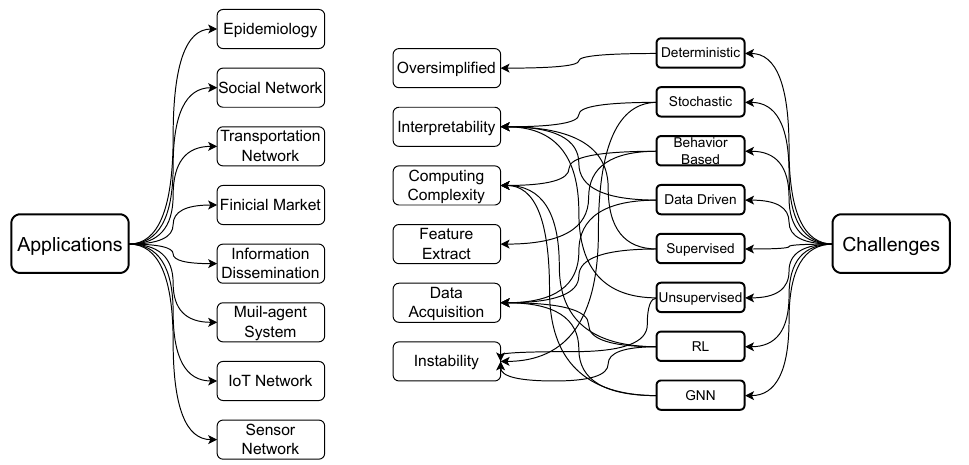}  
    \caption{Applications and Challenges}  
    \label{fig:applications_and_challenges_graph}  
\end{figure*}

\subsection{Graph Neural Network}
Based on the classification by training methods, Graph Neural Networks do not transcend the above three machine learning algorithms and cannot be categorized as a separate class \cite{xiao2021learning,rahmani2023graph,jiang2022graph}. However, considering the strong correlation between graph neural networks and network modeling, we will discuss them separately here.

Graph Neural Networks (GNNs) have a unique advantage in handling the information propagation characteristics within networks, particularly in capturing complex interactions and dependencies between nodes \cite{yu2020identifying,liu2024review}. 
GNN not only has a wide range of applications in network propagation problems \cite{yu2021self,cao2022mepognn,mahmud2021human,sivakumar2023enhancing,dong2023graph,li2023graph}, but can also be combined with other machine learning methods \cite{murphy2021deep,gangireddy2020unsupervised,wang2020risk} to enhance the overall performance of the model.

Assume we have a social network where individuals (nodes) are connected through relationships (edges). Our goal is to predict the coverage of a piece of information that starts propagating from a specific source node.

Let \( G = (V, E) \) represent a graph, where each node \( v \) has a feature vector \( x_v \). Using Graph Convolutional Networks (GCN) as an example, the hidden state \( h_v^{(l+1)} \) of each node at layer \( l+1 \) is updated as follows:

\begin{equation}
    h_v^{(l+1)} = \sigma\left( W^{(l)} \sum_{u \in \mathcal{N}(v) \cup \{v\}} \frac{h_u^{(l)}}{|\mathcal{N}(v)| + 1} + b^{(l)} \right)
\end{equation}

\noindent where \( \sigma \) is a non-linear activation function, \( W^{(l)} \) and \( b^{(l)} \) are learnable parameters, and \( \mathcal{N}(v) \) is the set of neighboring nodes of node \( v \).

For the information propagation task, a global readout layer can be set up to aggregate information from all nodes and predict the overall propagation impact across the network:

\begin{equation}
    \hat{y} = \text{softmax}\left( W^{\text{readout}} \sum_{v \in V} h_v^{(L)} \right)
\end{equation}

\noindent where \( L \) is the final layer, and \( W^{\text{readout}} \) is the parameter of the readout layer.

Compared to traditional deep learning and machine learning methods, the greatest uniqueness of GNNs in addressing network propagation problems lies in structured information processing. GNNs operate directly on graph structures, allowing them to naturally capture relationships and dependencies between nodes, something that is difficult to achieve with traditional neural networks or non-network-specific machine learning methods.


\section{Applications and Challenges}
Among the two main categories and their corresponding eight methods mentioned above, we can use eight-dimensional evaluation criteria to assess these methods, see Table \ref{table:evaluation}. Additionally, we will further introduce the applications and challenges of these methods.
The main advantages of deterministic models lie in their clear mathematical structure and predictability. Using differential or difference equations, they can clearly describe the behavior of a system. For example, the basic reproduction number (\(R_0\)) can effectively analyze the potential spread of infectious diseases. Parameters such as transmission rate (\(\beta\)) and recovery rate (\(\gamma\)) can be used to assess the speed and scope of virus or information diffusion in a network.
However, deterministic models assume that all system behavior is determined by initial conditions and parameters, without considering any randomness or external uncertainties. In practical applications, interactions between individuals often involve uncertainty, and transmission may have randomness, making deterministic models potentially overly simplistic and unable to accurately reflect real situations.
Moreover, in cases of large network scales or complex transmission mechanisms, analytical solutions are often difficult to obtain. In such cases, numerical simulations must be relied upon, increasing computational complexity. Many assumptions in deterministic models may not apply to real-world complex networks. For instance, the SIR model assumes that once a node recovers, it cannot be infected again. However, for certain infectious diseases (like influenza) or information spread, individuals may re-engage in the transmission process, which limits the applicability of the model.
Deterministic models, particularly the SIR model, are widely used in epidemiology. Health authorities use this model to predict the spread of diseases in different populations, such as analyzing the spread of COVID-19 and formulating intervention measures. In social networks, deterministic models can be used to simulate the spread of rumors, helping to understand how information propagates through different social connections, eventually reaching a stable state or disappearing.
Deterministic models are very useful in simplifying complex systems and providing precise analyses of their behavior, especially in application scenarios with clear transmission mechanisms. However, when dealing with real-world randomness and complexity, they need to be combined with other models (such as stochastic or data-driven models).

By introducing random variables, stochastic models can effectively handle the inevitable randomness and uncertainty in the real world. For example, user behavior in social networks or individual behavior in disease transmission is often unpredictable. Stochastic models account for these uncertainties, generating simulation results that more closely resemble reality.
Through analyzing long-term behaviors, such as the **stationary distribution** in random walk models, we can study the ultimate impact and speed of information dissemination in networks. This is crucial for understanding how network structures affect transmission, particularly in long-term prediction analyses of social networks or epidemic spread.
Compared to some complex deterministic models, stochastic models can, in certain cases, efficiently yield results through Monte Carlo simulations or other numerical methods of stochastic processes, without the need to solve differential or difference equations exactly. However, due to the introduction of randomness, even under the same initial conditions, each simulation may yield different results. This leads to relatively poorer interpretability and stability, making it difficult to derive definitive conclusions in practical applications. Conclusions often need to be drawn through extensive simulations and statistical analysis.
Although stochastic models can provide approximate solutions in some cases, in large-scale networks—especially when multiple iterations are involved—the computational cost can still be high. Multiple simulations are required to obtain statistically stable results. The accuracy of stochastic models depends on the choice of random variables and probability distributions. If parameters are chosen poorly, the model's results may be biased. Therefore, parameter adjustment and model calibration are crucial steps in applying stochastic models.
In social networks, information dissemination is often random and unpredictable. Random walk models are commonly used to analyze the process of information spreading from one node to others. For example, user behavior on Twitter, such as retweeting, can be viewed as a random process where users may randomly choose whether to forward information. By calculating the stationary distribution of different nodes, we can identify "influencers" in social networks—nodes that are most likely to be influenced by information over the long term.
In the case of virus transmission, the classic SIR model can incorporate stochasticity, simulating the probability of individuals getting infected or recovering after being exposed to the virus. For instance, in COVID-19 transmission research, stochastic models have been used to simulate the contact process and transmission risk among individuals. Through multiple simulations, the spread speed and range of the epidemic under different scenarios can be estimated.
In the field of transportation networks, particularly in studying traffic congestion and delay propagation, stochastic models are also useful. For instance, vehicle flow and route choices in a traffic network exhibit randomness, which can be modeled using random walk techniques to simulate dynamic changes in traffic flow. By simulating the random processes of different route choices, potential traffic bottlenecks can be predicted.

Agent-based models (ABMs) capture individual behavioral differences and decision-making processes by modeling each entity in a system as an independent agent. These models offer great flexibility, making them suitable for handling complex and dynamic systems.
ABMs allow researchers to adjust a wide range of parameters, such as adjacency relationships, behavioral rules, and decision thresholds, enabling flexible scenario simulations and experiments. By modifying these parameters, models can adapt to various scenarios, such as capturing different types of spread or interaction behaviors in traffic, economic, or social networks.
However, as the number of agents and the complexity of interactions increase, the demand for computational resources rises sharply. The state update for each agent typically requires considering the states of its neighbors and potential random events, leading to a computational complexity of \( O(n^2) \), where \( n \) is the number of agents. In practical applications, ABMs require extensive parameter calibration, including agent attributes, decision rules, and interaction weights. The performance of the model is directly influenced by the choice of these parameters, and when data is insufficient or prior knowledge is lacking, parameter tuning can be challenging.
ABMs involve substantial randomness and complex interactions during simulation. While they can capture the overall behavior of a system, they often lack intuitive explanations, making it difficult to extract clear causal relationships from the results, especially when studying specific mechanisms of transmission or behavior motivations.
In epidemic prevention, ABMs can simulate individual behavioral patterns during an outbreak. For example, by setting different behavioral rules (such as vaccination, isolation, or mask-wearing), the impact of these measures on disease transmission can be analyzed. Agents’ states can be defined as "susceptible," "infected," or "recovered" (following the SIR model), and their state updates are influenced by the infection status of surrounding individuals and personal protective behaviors.
In intelligent transportation systems, ABMs can simulate vehicle or pedestrian behavior. Each agent represents a vehicle or pedestrian, and the model can adjust route choices and speed based on traffic signals, the behavior of other vehicles, and road conditions. By adjusting the agents’ decision-making rules, the effects of different traffic management measures on traffic flow can be analyzed.
In financial markets, ABMs can simulate investors’ trading decisions. Each agent represents an investor, and their behavior rules can be based on market information (such as stock prices or news events) and the actions of other investors. For instance, agents may update their investment strategies based on the buy/sell decisions of their neighbors, thereby simulating herd behavior or price bubbles in the market.

Data-driven models can continuously update and optimize themselves based on new data. By dynamically adjusting propagation parameters, these models are better suited to adapt to real-time changes in network environments and dissemination patterns. They are particularly effective at capturing nonlinear characteristics in complex networks, often providing more accurate predictions of propagation than models based solely on theoretical derivations.
The accuracy and effectiveness of data-driven models depend heavily on the quality of the training data. If the data contains noise, is incomplete, or biased, the model’s performance and predictive ability may be compromised. Additionally, when there is insufficient historical data, the model may struggle to reliably learn accurate propagation parameters.
In epidemic prevention, data-driven models can use historical infection data to learn the transmission rate \( \beta(t) \) and recovery rate \( \gamma(t) \) of a disease. These models can dynamically adjust and predict the future spread of the epidemic and the number of infections, aiding governments in formulating control policies. 
On social media platforms like Twitter or WeChat, data-driven models can learn information dissemination patterns based on historical data on retweets, comments, and likes. The model can help predict the future spread of a message and identify key nodes that will influence its dissemination.

Deep learning-based supervised learning methods offer unique advantages and challenges when addressing the issue of network information dissemination. These models can automatically learn complex features from input data, reducing reliance on feature engineering required in traditional machine learning methods. By utilizing deep neural networks, models can capture nonlinear relationships and higher-order interactions in the data.
These models excel at handling large-scale datasets, making them particularly effective in identifying complex propagation patterns in large-scale social or transmission networks. For example, by inputting information about nodes and their neighbors, deep learning models can effectively model how information spreads from one node to another. Additionally, they can be applied to various types of data, including images, text, and structured data. Even though the network structure itself may be complex, deep learning models can handle this effectively through feature representations, such as node attribute matrices.
However, the performance of supervised learning models depends on the availability of sufficient and high-quality labeled data. For network dissemination problems, obtaining enough labeled data (such as records of propagation events) can be challenging, especially in large-scale networks.
Deep learning models, especially deep neural networks, are often regarded as "black box models," meaning that their decision-making process is difficult to interpret. This can be a challenge in network dissemination problems, as researchers or decision-makers may want to understand which features or nodes are influencing the spread process.
While deep learning models have strong fitting capabilities, they may overfit when data is limited or imbalanced, performing well on training data but generalizing poorly to new data. Therefore, regularization techniques and cross-validation are crucial for mitigating overfitting.
On social media platforms, deep learning can be used to predict the reach and speed of message or news dissemination. By inputting user characteristics, message content, and social relationships, models can predict which messages are likely to go viral.
In epidemiology, deep learning can be used to predict the transmission paths and scope of disease spread. Using data such as geographic location, population density, and contact networks, models can help health departments identify areas most susceptible to outbreaks, optimizing vaccine distribution or quarantine policies.
In marketing campaigns, companies can use deep learning models to predict the effectiveness of ads, promotions, or brand messaging. By analyzing consumer behavior and social network structures, models can help businesses identify which nodes hold the most influence, thereby optimizing marketing strategies.

Unsupervised learning does not require pre-labeled data, which offers significant advantages in practical applications. Many network datasets are vast and unlabeled, and unsupervised methods like Generative Adversarial Networks (GANs) can fully utilize this unlabeled data for model training. This makes such methods particularly useful in scenarios where data acquisition costs are high or data labeling is incomplete.
Deep learning methods, especially GANs, can automatically discover complex patterns and features within the data. By generating realistic data samples, GANs can capture the underlying distribution of the data, effectively simulating complex real-world information dissemination processes.
However, in addition to inheriting most of the challenges of neural network-based supervised learning, unsupervised network training, such as in GANs, often faces instability. The adversarial relationship between the generator and the discriminator can lead to unstable training processes, making it prone to **mode collapse**—a situation where the generator outputs a limited range of sample types and fails to capture the full data distribution. 
Moreover, ensuring convergence can be difficult, and careful hyperparameter tuning is required to stabilize training and achieve optimal results.

Unlike traditional methods, reinforcement learning (RL) can autonomously learn optimal strategies for information dissemination through trial-and-error interactions, without relying on pre-built models. This method can also optimize multiple objectives simultaneously, such as maximizing information coverage, minimizing transmission costs, or controlling propagation speed. By designing appropriate reward functions, deep reinforcement learning (DRL) can balance these goals effectively.
The training process for RL requires a significant number of interactions and iterations, making training computationally expensive. Additionally, the effectiveness of RL is highly dependent on the design of the reward function. If the reward function is poorly designed, the model may learn suboptimal or ineffective strategies. Furthermore, balancing multiple objectives when designing the reward function can be challenging.
RL is widely used in multi-robot systems, particularly when multiple robots need to collaborate to complete tasks. DRL can help robots learn how to allocate tasks and share information to maximize the overall efficiency of the system. For instance, in disaster rescue operations, multiple robots must cooperate in search-and-rescue missions, and DRL can optimize their collaborative strategies.
In the Internet of Things (IoT) networks, DRL can help optimize data transmission between sensor nodes, ensuring that critical data is quickly and accurately transmitted to central nodes or gateways while minimizing energy consumption. This is essential for data management in smart cities or smart home systems.
In intelligent transportation systems, DRL can dynamically adjust traffic signal control strategies and optimize vehicle route planning based on real-time data feedback, reducing traffic congestion and improving road utilization. This application effectively addresses the dynamic changes in large-scale, complex traffic networks.

Graph Neural Networks (GNNs) operate directly on graph structures, automatically capturing relationships and dependencies between nodes, which is crucial for solving propagation problems in complex networks. Compared to traditional neural networks, GNNs are better suited for handling structured data consisting of nodes and edges, effectively learning both local and global information about nodes.
GNNs can be applied independently or combined with other machine learning methods, such as supervised learning, unsupervised learning, or reinforcement learning, to enhance overall model performance. For instance, combining GNN with reinforcement learning can help discover optimal propagation strategies in networks.
However, GNNs' computational complexity increases significantly with graph size. Specifically, calculating the features for each node requires aggregating the features of all its neighbors, which can quickly lead to high computational and memory consumption, especially when dealing with large-scale graphs.
GNN training requires large amounts of labeled data, particularly in supervised learning settings, and collecting high-quality labeled data can be costly. Additionally, GNNs have a large parameter space, making them prone to overfitting, thus requiring carefully designed regularization strategies and model tuning.
While GNNs perform well in handling static graph structures, they have limited adaptability to dynamic networks that change over time.
GNNs can effectively model disease transmission pathways in social or population networks. For instance, in modeling the spread of infectious diseases like COVID-19, GNNs can predict the scope and speed of the outbreak, assisting in the formulation of control measures.
In recommendation systems, users and items can be represented as a graph, and GNNs can capture the interactions between users and items, predicting potential interests and improving recommendation accuracy. For example, GNNs can analyze users' purchasing behavior and the influence of their social networks to recommend products or content that align with their interests.
In intelligent transportation systems, GNNs can be used to predict traffic flow and optimize route planning. Nodes in the traffic network represent intersections or traffic facilities, and edges represent road segments. By modeling traffic networks with GNNs, traffic management systems can predict congestion, optimize traffic signal control, and enhance overall traffic flow efficiency.

A summary of the applications and challenges discussed above can be found in Figure (\ref{fig: applications and challenges graph}).

\section{Conclusion}
In this article, we investigate fundamental approaches that can be applied to network propagation problems. We classify these approaches into two types: model-based and model-free methods.
For model-based methods, we begin with the most basic deterministic model approach and move on to introducing stochastic models, behavioral models, and data-driven models.
For model-free methods, we mainly introduce approaches that can be based on neural networks and deep learning, primarily divided into three categories: supervised learning, unsupervised learning, and reinforcement learning. Additionally, we also discuss a special type of neural network method, the application of graph neural networks in network propagation problems.

This survey provides a clear problem-solving framework for future researchers working on network propagation issues. In the article, we also discuss the advantages and limitations of each method. Each method can be used independently or in combination, depending on the specific situation.
Of course, due to space limitations, not all methods are listed in detail, but this does not hinder the article from offering researchers a unique perspective.

In the future, especially for dynamic network propagation problems, the combination of these methods and the convergence of their underlying philosophies will undoubtedly lead to new research directions and innovative approaches. With the development of technology and the emergence of new computational capabilities and data analysis tools, future researchers will be able to leverage these interdisciplinary methods to propose more refined and intelligent solutions.

\end{document}